\documentclass{article}
\usepackage[utf8]{inputenc}
\usepackage{amsmath}
\usepackage{amssymb}
\usepackage{amsfonts}
\usepackage{esvect}
\usepackage{cite}
\usepackage{physics}
\usepackage{mathrsfs}
\usepackage{authblk}
\usepackage{geometry}
 
\usepackage{abstract}
\usepackage{xcolor}
\title{\bf Quadratic algebras as commutants of algebraic Hamiltonians in the enveloping algebra of Schr\"odinger algebras  }
\author{\large Rutwig Campoamor-Stursberg$^{1}$\footnote{rutwig@ucm.es}, 
Ian Marquette$^{2}$ \footnote{i.marquette@uq.edu.au} }
\affil{$^{1}$ Instituto de Matem\'atica Interdisciplinar and Dpto. Geometr\'{\i}a y
Topolog\'{\i}a,
UCM,\\E-28040 Madrid, Spain}
\affil{$^{2}$ School of Mathematics and Physics, The University of Queensland \\ Brisbane, QLD 4072, Australia}

\usepackage[colorlinks=true,urlcolor=black]{hyperref}

\begin{document}

\maketitle
\begin{abstract}
We discuss a procedure to determine finite sets $\mathcal{M}$ within the commutant of an algebraic Hamiltonian in the enveloping algebra of a Lie algebra $\mathfrak{g}$ such that their generators define a quadratic algebra. Although independent from any realization of Lie algebras by differential operators, the method is partially based on an analytical approach, and uses the coadjoint representation of the Lie algebra $\mathfrak{g}$. The procedure, valid for non-semisimple algebras, is tested for the centrally extended Schr\"odinger algebras $\widehat{S}(n)$ for various different choices of algebraic Hamiltonian. For the so-called extended Cartan solvable case, it is shown how the existence of minimal quadratic algebras can be inferred without explicitly manipulating the enveloping algebra. 
\end{abstract}

\section{Introduction}

 The classification and hierarchization problem of superintegrable systems has shown that the notion of symmetry can be interpreted far beyond the classical approach using Lie algebras typically appearing in the context of  integrals of the motion or spectrum-generating algebras, incorporating other types of algebraic structures such as quadratic algebras \cite{Fre91,Kress07,Mill13}. These algebras appear naturally when analyzing the commutation relations of conserved quantities, and provide an insightful connection with various other mathematical tools, such as the Askey scheme of orthogonal polynomials, the recoupling coefficient formalism, the coalgebra method or the Clebsch-Gordan problem \cite{Yat18,Lati21}. These extended objects, although still incompletely exploited, have been shown to be of crucial importance in order to establish a hierarchy of superintegrable systems and their mutual connection. A powerful illustration of this fact is given by the scheme relating the twelve classes of superintegrable systems on 2D conformally flat space by means of contractions of their corresponding quadratic algebras \cite{Mill13}. This important result suggests that an analogous behaviour should be expected for higher-dimensional systems, a problem that motivates further research. In addition, deep relations with generalized special functions and algebras introduced and studied in a slightly different context, such as the Bannai-Ito and Racah algebras \cite{Lati21}, have been profitably applied to the superintegrability classification problem. A certain constraint is however  given by the fact that many of these approaches are dependent on explicit realizations of Lie algebras in terms of differential operators, which conditionally implies some consequences that emerge not from the underlying symmetry algebra, but from the explicit choice of the realization. In this context, some purely algebraic schemes have been proposed recently, in order to circumvent this restriction and better understand these observed pathologies (see e.g. \cite{Mill13,Lati21,Yi18} and references therein). In this approach, the Hamiltonian of the system is interpreted algebraically in terms of some Lie algebra $\mathfrak{g}$, with the corresponding integrals of the motion constructed in the corresponding enveloping algebra $\mathcal{U}(\mathfrak{g})$. This ansatz was used in previous papers, where some constructions of algebraic Hamiltonians and quadratic algebras related to the reductive Lie algebras $\mathfrak{gl}(3)$ and $\mathfrak{su}(3)$ have been considered \cite{cam21,cor21}. 
 
 \medskip
 \noindent The purpose of this paper is twofold. One the one hand, we intend to develop a setting for the construction of quadratic algebras on purely algebraic grounds more systematically, without referring to an explicit realization in terms of differential operators, but without neglecting the deep relation with the analytical approach, more precisely, the invariant theory associated to the coadjoint representation of Lie algebras that emerges from the identification of Casimir operators with the commutant of a Lie algebra in its enveloping algebra \cite{Gel,Dix}. This is explicitly used to derive some polynomials that commute with each of the generators appearing in the Hamiltonian $\mathcal{H}$, to which eventually additional polynomials are added to generate a suitable quadratic algebra contained in the commutant  $C_{\mathcal{U}(\mathfrak{g})}(\mathcal{H})$. This leads to a notion of minimality of quadratic algebras, defined in terms of integrity bases of certain systems of partial differential equations (PDEs in short). The problem is formulated and studied in the context of the centrally extended Schr\"odinger algebra $\widehat{S}(n)$, i.e., the maximal invariance group of the free Schr\"odinger  equation \cite{Ni72,Fei04}, used in the study of  Appell systems and quantum groups, and recently generalized to the class of conformal Galilei algebras \cite{Ai13,Ai15,cam19}. Various algebraic Hamiltonians related to distinguished subalgebras of $\widehat{S}(n)$ are analyzed in this context, showing that for this class of Lie algebras, a systematic prescription to determine minimal quadratic for certain choices of algebraic Hamiltonians can be formulated.

\subsection{Commutants in $ \mathcal{ U}(\mathfrak{g})$ and quadratic algebras}

\noindent Given a Lie algebra $\mathfrak{g}$ of dimension $n$ and basis $\left\{X_1,\dots ,X_n\right\}$, we denote the symmetric algebra associated to $\mathfrak{g}$ as $S(\mathfrak{g})$, which can usually be identified with a polynomial ring in $n$ commuting variables   $\left\{x_1,\dots ,x_n\right\}$ \cite{AMA75}. By means of the Poincar\'e--Birkhoff--Witt theorem, the unit $1$ and the monomials $X_{i_1}^{\alpha_1}\cdots X_{i_s }^{\alpha_s}$ with $1 \leq i_1 \leq  \cdots \leq i_s\leq n $, form a basis of $ \mathcal{ U}(\mathfrak{g})$. For a fixed number $k$, we denote by $\mathcal{U}_{(k)}(\mathfrak{g})$ the linear subspace generated by the monomials $X_1^{a_1}\dots X_n^{a_n}$ satisfying the constraint $a_1+a_2+\dots +a_n\leq k$. This allows to define the degree ${\rm deg}(U)$ of an element $U\in\mathcal{U}(\mathfrak{g})$ as  ${\rm deg}(U)={\rm inf}\left\{k\;|\; U\in \mathcal{U}_{(k)}(\mathfrak{g})\right\}$ and the natural filtration $\mathcal{U}_{(k)}(\mathfrak{g})\subset \mathcal{U}_{(k+1)}(\mathfrak{g})$ for $k\geq 0$. It follows that 
\begin{equation}
\left[\mathcal{U}_{(j)}(\mathfrak{g}),\mathcal{U}_{(k)}(\mathfrak{g})\right]\subset \mathcal{U}_{(j+k)}(\mathfrak{g}),\quad j,k\geq 0.\label{fil1}
\end{equation}

\medskip
\noindent The adjoint action of the Lie algebra $\mathfrak{g}$ on the symmetric algebra $S(\mathfrak{g})$ and the enveloping algebra $\mathcal{U}(\mathfrak{g})$ is given, respectively, by the formulae for the basis  $\left\{X_1,\dots ,X_n\right\}$:
\begin{equation}\label{adja}
\begin{array}[c]{rl}
P\left(x_1,\dots ,x_n\right)\in S(\mathfrak{g})\mapsto &\displaystyle  \widehat{X}_i(P)=C_{ij}^{k} x_k \frac{\partial}{\partial x_j}\in S(\mathfrak{g}),\\
U\in \mathcal{U}(\mathfrak{g}) \mapsto & \left[X_i,U\right]= X_i U-U X_i\in\mathcal{U}(\mathfrak{g}) 
\end{array}
\end{equation}
It is well known that by means of the symmetrization map
\begin{equation}\label{syma}
\Phi\left(x_{i_1}x_{i_2}\dots x_{i_s}\right)=\frac{1}{s!} \sum_{\sigma\in\Sigma_{s}} X_{i_{\sigma(1)}}X_{i_{\sigma(2)}}\dots X_{i_{\sigma(s)}},
\end{equation}
with $\Sigma_s$ the symmetric group in $s$ letters, we can construct a linear isomorphism $\Phi:S(\mathfrak{g})\rightarrow \mathcal{U}(\mathfrak{g})$ that commutes with the adjoint action, i.e., such that $\Phi\left(\widehat{X}(P)\right)=\left[X,\Phi(P)\right]$ for all $X\in\mathfrak{g},\; P\in S(\mathfrak{g})$. In particular, if $S^{(k)}(\mathfrak{g})$ denotes the space of homogeneous polynomials of degree $k$, then $\mathcal{U}^{(k)}(\mathfrak{g})=\Phi\left(S^{(k)}(\mathfrak{g})\right)$ implies that $\mathcal{U}_{k}(\mathfrak{g})=\sum_{\ell=0}^{k} \mathcal{U}^{(\ell)}(\mathfrak{g})$ and for arbitrary elements $Q_1\in S^{(p)}(\mathfrak{g})$, $Q_2\in S^{(q)}(\mathfrak{g})$ following relations hold: 
\begin{gather}
{\rm deg}\left(\Phi(Q_1Q_2)\right)= p+q,\quad \Phi(Q_1Q_2)-\Phi(Q_1)\Phi(Q_2)\in \mathcal{U}_{(p+q-1)}(\mathfrak{g}).\label{grad}
\end{gather}

 \noindent As invariants in $S(\mathfrak{g})$ and ${\cal U}(\mathfrak{g})$ under the adjoint action of $\mathfrak{g}$  we understand the following sets:
\begin{equation}
\begin{split}
\mathcal{U}(\mathfrak{g})^{I} = \left\{ U\in\mathcal{U}(\mathfrak{g})\quad |\quad \left[X_{i},U\right]=0,\;  1\leq i\leq n\right\},\\
S(\mathfrak{g})^{I} =  \left\{ P\in S(\mathfrak{g})\quad |\quad \widehat{X}_{i}(P)=0,\;  1\leq i\leq n\right\}.\label{INVS1}
\end{split}
\end{equation}
Clearly these spaces are linearly isomorphic, and as Abelian algebras they are moreover algebraically isomorphic, although in general, this algebraic isomorphism does not coincide with the symmetrization map \cite{Dix}. 

\medskip
\noindent  The commutant $C_{\mathcal{U}(\mathfrak{g})}(A)$ of an element $A\in\mathcal{U}(\mathfrak{g})$ is defined  as the set of elements in $\mathcal{U}(\mathfrak{g})$ that can be permuted with $A$, that is, 
\begin{equation}
C_{\mathcal{U}(\mathfrak{g})}(A)=\left\{ U\in\mathcal{U}(\mathfrak{g})\; |\; [A,U]=0\right\}.\label{comm}
\end{equation}
 
\medskip
\noindent Now suppose that $Y\in\mathfrak{g}$ is an arbitrary element. In order to determine the commutant $C_{\mathcal{U}(\mathfrak{g})}(Y)$ of $Y$ in the enveloping algebra, we can use that fact that the symmetric algebra $S(\mathfrak{g})$ can be identified with the algebra of polynomial functions over $\mathfrak{g}^{*}$ \cite{Dix1}, hence enabling us to consider the coadjoint representation of $\mathfrak{g}$ (see e.g. \cite{Tro}) and then using the symmetrization map (\ref{syma}) to obtain those elements of $C_{\mathcal{U}(\mathfrak{g})}(Y)$ having minimal degree. As $Y$ is expressible as $Y= a^{i} X_i$ for certain constants $a^{i}$ (over a basis $\left\{X_1,\dots ,X_n\right\}$ of $\mathfrak{g}$), the polynomial solutions of the partial differential equation (see (\ref{adja}))
\begin{equation}
\widehat{Y}(\varphi)= a^iC_{ij}^k x_k \frac{\partial \varphi}{\partial x_j}= a^{i}\widehat{X}_i(\varphi)=0 \label{inv2}
\end{equation}
correspond, after symmetrization, to elements in $C_{\mathcal{U}(\mathfrak{g})}(Y)$. The same argument can be applied to a subalgebra $\mathfrak{h}$, where we consider the system of PDEs associated to the generators of the subalgebra. 
As these are homogeneous linear first-order systems of PDEs, we can always find an integrity basis $\mathcal{I}$ for the solutions \cite{Kam}, showing that the commutant $\mathcal{U}_{\mathfrak{h}}(\mathfrak{g})$ is actually generated by the symmetrized images in $\mathcal{U}(\mathfrak{g})$.\footnote{If a certain system does not admit an integrity basis generated by polynomials, such as happens for instance for certain solvable Lie algebras, then the nonpolynomial  elements must be skipped.}   

\medskip
\noindent For elements $U\in\mathcal{U}(\mathfrak{g})$ of degree $d\geq 2$, hence not belonging to $\mathfrak{g}$, the situation is more complicated from the computational point of view. Although commuting polynomials can still be analyzed in terms of differential operators, it must be taken into account that the differential operator $\widehat{U}$ associated to $U$ is of order $d$, making manipulations as cumbersome as a direct inspection of the enveloping algebra. However, some information may still be obtained using the analytical approach based on the coadjoint representation. To this extent, let be $X_{i_1},\dots ,X_{i_k}$ the generators of $\mathfrak{g}$ such that $U$ can be written as a polynomial $P(X_{i_1},\dots ,X_{i_k})$. If now $\varphi(x_1,\dots ,x_n)$ is a common solution to the system of PDEs $\widehat{X}_{i_{s}}(\varphi)=0$ for $1\leq s\leq k$, it is straightforward to verify that the symmetrization $\Phi(\varphi)$ belongs to the commutant. If $U$ is itself a solution of the preceding system, then all commuting elements can be found analytically. Otherwise, the commutant contains additional operators that do not correspond to integrals of the system, such as $U+\varphi$.\footnote{In this context it is relevant to recall that, as already observed in \cite{cam21}, for a given realization of $\mathfrak{g}$ in terms of differential operators nonequivalent to that obtained from the coadjoint representation, there may be differential operators that commute with a Hamiltonian $\mathcal{H}$ in the realization, but that do not correspond to commuting elements in the enveloping algebra. }   

\medskip
\noindent As an example to illustrate this procedure, consider the solvable three-dimensional Lie algebra $\mathfrak{r}$ with brackets $[X_3,X_1]=X_2$, $[X_3,X_2]=-X_1$, $[X_2,X_3]=0$ and the Hamiltonian $\mathcal{H}_0=X_3+X_1^2$. It is immediate to verify that the system of PDEs $\widehat{X}_1(\varphi)=0,\; \widehat{X}_3(\varphi)=0$ associated to the generators $X_1,X_3$ admits as general solution a function of the polynomial $I_2=x_1^2+x_2^2$, leading after applycation of (\ref{syma})  to the element $A_1=X_1^2+X_2^2\in\mathcal{U}(\mathfrak{r})$. As  $\mathcal{H}_0$ does not commute with $X_1$, $A_1$ is an independent element in the commutant, and the difference $A_2=A_1-\mathcal{H}_0=X_2^2-X_3$  also belongs to $C_{\mathcal{U}(\mathfrak{g})}(\mathcal{H}_0)$. It can actually be verified that any element in the commutant is a polynomial in $A_1$ and $A_2$, implying that $C_{\mathcal{U}(\mathfrak{g})}(\mathcal{H}_0)$ is an Abelian algebra isomorphic to $\mathbb{R}\left[A_1,A_2\right]$.   
 
\medskip
\noindent Given an algebraic Hamiltonian $\mathcal{H}$ defined in terms of the generators of a Lie algebra $\mathfrak{g}$ and its commutant $C_{\mathcal{U}(\mathfrak{g})}(\mathcal{H})$ in the enveloping algebra, our interest will not be focused on the particular structure of the commutant, but on the problem whether we can find an appropriate subset $\mathcal{M}$ such that its elements define a non-Abelian quadratic algebra with respect to the usual commutator in $\mathcal{U}(\mathfrak{g})$. Certainly, the existence of such quadratic subalgebras, if any at all, is by no means uniquely determined, so that certain restrictions to the problem have to be imposed. The correspondence between 
$\mathcal{U}(\mathfrak{g})^{I} $ and $S(\mathfrak{g})^{I}$ suggests to make use of the analytical approach, even if the precise decomposition of commutators of polynomials in $C_{\mathcal{U}(\mathfrak{g})}(\mathcal{H})$ must generally be computed algebraically, with the help of symbolic computer packages.

\medskip
\noindent The general procedure to find quadratic algebras that will be used in the following can be summarized as follows: 
Suppose that  the Hamiltonian $\mathcal{H}_0$ is given as a polynomial $P(X_{i_1},\dots ,X_{i_k})$  in terms 
of the generators of the Lie algebra $\mathfrak{g}$ (with basis $\left\{X_1,\dots ,X_n\right\}$). For each generator $X_{i_s}$ intervening in the expression of $\mathcal{H}_0$, we 
consider the differential operator $\widehat{X}_{i_j}$ given in (\ref{adja}) and solve the system of PDEs 
\begin{equation}
\widehat{X}_{i_1}\left(F(x_1,\dots ,x_n)\right)=0,\quad 1\leq j\leq k.\label{sys}
\end{equation}
There are always $n-r_0$ independent solutions, where $r_0$ denotes the generic rank of the system. Depending on the structure of the Lie algebra $\mathfrak{g}$, the system may admit an integrity basis $\mathcal{I}$ formed by polynomials. If this is not the case, then we consider a maximal set of (functionally) independent polynomial solutions $\left\{Q_1,\dots ,Q_p\right\}$ of (\ref{sys}), where $p\leq n-r_0$. After symmetrization using the map (\ref{syma}), we obtain elements $M_j=\Phi(Q_j)$ of degree $d_j$ in the commutant. There may be additional elements $M_{j}^{\prime}$ in $C_{\mathcal{U}(\mathfrak{g})}(\mathcal{H})$ having degrees $d\leq d_j$ that do not commute with all the generators $X_{i_s}$, as in general the Hamiltonian $\mathcal{H}_0$ itself is not a solution of (\ref{sys}). Starting from a set of polynomials $\mathcal{S}=\left\{\mathcal{H}_0,M_1,\dots ,M_p,M_1^{\prime},\dots ,M_q^\prime\right\}$ of degree $d\leq \max \left\{d_j\;|\; 1\leq j\leq p\right\}$, we analyze whether their commutators can be expressed as quadratic polynomials in the generators, and eventually discard or add new elements to $\mathcal{S}$  until a quadratic algebra $\mathcal{A}$ of minimal dimension is found. By ``minimal" we mean that excluding one or more elements of $\mathcal{A}$ results in the commutators being either linear or not expressible in terms of the generators.

\section{The one-dimensional Schr\"odinger algebra $\widehat{S}(1)$}

The (centrally extended) Schr\"odinger algebra $\widehat{S}(1)$ in (1+1)-dimensions is isomorphic to the semidirect sum $\mathfrak{sl}(2)\overrightarrow{\oplus}_{D_{\frac{1}{2}}\oplus D_0}\mathfrak{h}_1$, where $\mathfrak{h}_1$ denotes the three-dimensional Heisenberg algebra. 
Over the basis $\left\{H,E,F,P_0,P_1,m\right\}$, the commutators are  
\begin{equation}\label{S1}
\begin{array}[c]{llll}
\left[H,E\right]=2 E, & [H,F]=-2 F,& [H,P_{0}]=P_{0}, & [H,P_{1}]=-P_{1},\\
\left[E,F\right]=H, & [E,P_{0}]=0, &  [E,P_{1}]=P_{0}, &  [F,P_{0}]=P_{1},\\
\left[F,P_{1}\right]=0, &  [P_{1},P_{0}]=m. & &
\end{array}
\end{equation}
This Lie algebra has two Casimir operators, one being the central charge $m$, while the second is given by the third-order polynomial
\begin{equation}
{\bf C}_0=\frac{m}{2}(H-H^2)-2mEF+P_1P_0-FP_0^2+P_1HP_0+P_1^2E \label{casim}
\end{equation}
By (\ref{INVS1}), the set of invariants $\mathcal{U}(\mathfrak{g})^{I}$ is generated by ${\bf C}_0$ and $m$. 

\medskip
\noindent 
As a first trivial albeit illustrative example of a quadratic algebra contained in a commutant, we shall construct the commutant of the Cartan generator $\mathcal{H}_0=H$
of  $\widehat{S}(1)$ in the enveloping algebra $\mathcal{U}( \widehat{S}(1))$. It is obvious from (\ref{S1}) that the 
only linear polynomials commuting with $\mathcal{H}_0$ are $H$ and the central charge $m$, that generate an Abelian algebra. In the 
following, as $m$ commutes with any generator and thus with any polynomial, we will consider $m$ as a scalar or parameter, 
hence it will be discarded in the explicit analysis of commutants. If $\left\{h,e,f,p_0,p_1,m\right\}$ denote the commuting variables 
in $\widehat{S}(1)^{\ast}$, the commutant of $H$ is obtained 
symmetrizing the polynomials satisfying the PDE
\begin{equation}
\widehat{H}(\varphi)= 2e \frac{\partial \varphi}{\partial e}-2f \frac{\partial \varphi}{\partial f}+p_0\frac{\partial \varphi}{\partial p_0}-p_1\frac{\partial \varphi}{\partial p_1}=0. \label{inca}
\end{equation} 
It is easy to verify that we can consider the five polynomials
\begin{gather}
Q_{1}=h ,\quad Q_{2}=p_{1} p_{0} ,\quad Q_{3}=e f, \quad Q_{4}=p_{1}^{2} e,\quad Q_5= m
\end{gather}
as an integrity basis for the solutions of (\ref{inca}). Ignoring $m$ for the reasons already mentioned, symmetrizing the remaining polynomials and simplifying the resulting expressions, we find that the commutant $\mathcal{U}_{H_0}(\widehat{S}(1))$ is generated by the elements 
\begin{gather}
A_{1}=H ,\quad A_{2}=P_{1} P_{0} ,\quad A_{3}=E F, \quad A_{4}=P_{1}^{2} E. \label{inta3}
\end{gather}
 However, as $\left[A_{2},A_{3}\right]=-A_{4}-F P_{0}^{2}$, which cannot be expressed quadratically in terms of the elements in (\ref{inta3}), we need to adjoin additional (algebraically dependent) polynomials in order to get a quadratic algebra. Considering the previous generators and the third-order polynomials 
\begin{gather}
A_{5}=P_{1} H P_{0},  \quad  A_{6}= E H F ,\quad A_{7}= F P_{0}^{2},\label{inta4}
\end{gather}
besides the trivial relations $ [A_{1},A_{i}]=0]$ for $i=1,\dots ,7$, 
these elements generate a quadratic algebra with commutators
\begin{equation}
\begin{array}[c]{llll}
\left[A_{2},A_{3}\right]=-A_{4}-A_{7} ,\quad  [A_{2},A_{4}]=-m A_{2}-A_{2}^{2}- 2 m A_{4},\quad \left[ A_{2},A_{5}\right]=0 , \quad  [A_{2},A_{6}]=(2 - A_{1}) A_{4} + (2 - A_{1}) A_{7},\nonumber\\
\left[A_{2},A_{7}\right]= - m A_{1} - A_{2}^{2} + 2 m A_{7} ,\quad  [A_{3},A_{4}]= m A_{1} - 2 A_{1} A_{2} - m A_{3} + 2 A_{2} A_{3} - A_{1} A_{4},\quad [A_{3}, A_{6}]=0,\nonumber\\
\left[ A_{3},A_{5}\right]= A_{4} + A_{1} A_{4} + A_{7} + A_{1} A_{7},\quad \left[ A_{3}, A_{7}\right]= m A_{1} - 2 A_{1}A_{2} - m A_{3} + 2 A_{2} A_{3} + 2 A_{4} + 2 A_{7} + A_{1} A_{7}, \nonumber\\
\left[ A_{4}, A_{5}\right]= m A_{2} + m A_{1} A_{2} +2 m A_{4} + 2 m A_{1} A_{4} + A_{2} A_{5}, \quad \left [ A_{4}, A_{6}\right]=0 ,\nonumber\\
\left[A_{4},A_{7}\right]= 2 m^{2} A_{1}+2 m A_{2} - 3 m A_{1} A_{2} + A_{2}^{2} - 2 m ^{2} A_{3} + 4 m A_{2} A_{3} + 4 m A_{4} + A_{2} A_{5}, \nonumber \\
\left[ A_{5}, A_{6}\right]=0 ,\quad [A_{5},A_{7}]=-m A_{2} -m A_{1} A_{2} - A_{2} A_{5} +2 m A_{7} + 2 m A_{1} A_{7}, \nonumber\\
\left[ A_{6}, A_{7}\right]= -2 m A_1  + m A_1^2 + 4 A_1 A_2 -2 A_1^2 A_2 +2 m A_3 - m A_1 A_3- 4 A_2 A_3 - 4 A_2 +2 A_1 A_2 A_3  \nonumber\\
 - 4 A_4 +2 A_1 A_4 -4 A_7 + A_1^2 A_7 \nonumber
 \end{array}\label{KLA}
\end{equation}

From the construction, it follows that the quadratic algebra is minimal among those that contain the generating set (\ref{inta3}). 
 As the noncentral Casimir operator ${\bf C}_0$ of $ \widehat{S}(1)$ commutes with all generators, a generic Hamiltonian $\mathcal{H}$ can be chosen as
$ \mathcal{H}= \lambda\mathcal{H}_0 + \mu{\bf C}_0 = \lambda H  + \mu {\bf C}_0 $, without altering the structure of the preceding quadratic algebra generated by the $A_i$. An explicit expression for $H$ can be obtained by means of the realizations as differential operators on the real plane given by 
\begin{equation}\label{REA1}
\begin{split}
H= t^2 \partial_t + t x_1 \partial_{x_1} + \frac{1}{2} t + \frac{1}{2}x_1^2 ,& \quad  E=- \partial_t,\quad  F= t^2 \partial_t +t x_1 \partial_{x_1}+ \frac{1}{2} t +\frac{1}{2}m x_1^2 \\
P_0=x_1 ,\quad P_1=-m x_1 - t \partial_{x_1}. &
\end{split}
\end{equation}

\subsection{Commutants associated to generators of subalgebras }

In this section we analyze more general commutants in the enveloping algebra of $\widehat{S}(1)$  that extend the construction of quadratic algebras. Three cases will be considered, called respectively the Borel, the extended Borel and the extended Cartan solvable algebra case. In particular, the last two show cases give rise to quadratic algebras of low dimension that can be easily extended adjoining additional polynomials of higher order. 

\subsubsection{Borel Case}

\noindent After the Cartan subalgebra, the Borel subalgebra $\mathfrak{b}$ of $\mathfrak{sl}(2,\mathbb{R})$ is the simplest choice for analyzing commutants in the enveloping algebra. As starting Hamiltonian we consider a parameterized linear combination of the Borel generators, i.e., $\mathcal{H}_0= H  + \lambda E$, where $\lambda\in\mathbb{R}-\{0\}$. As  $\mathcal{H}_0\in  \widehat{S}(1)$, we can first search for an integrity basis using the coadjoint representation. The differential equation to be considered in this case is given by 
\begin{equation}
-2\lambda e  \frac{\partial \varphi}{\partial h}+2 e \frac{\partial \varphi}{\partial e}+( \lambda h-2f)\frac{\partial \varphi}{\partial f}+p_0\frac{\partial \varphi}{\partial p_0}+( \lambda p_0-p_1)\frac{\partial \varphi}{\partial p_1}=0.\label{scsa}
\end{equation}
One possible integrity basis given by polynomials of minimal degree is given by 
\begin{equation}
Q_1=\lambda e+h,\quad Q_2=ef+\frac{1}{4}h^2,\quad Q_3=p_0^2-2\frac{1}{\lambda}p_0p_1,\quad Q_4=ep_0^2-\frac{4}{\lambda^2}fp_0^2+\frac{2}{\lambda}hp_0^2,\quad Q_5=m.\label{inta4}
\end{equation}
Symmetrizing these elements and considering $m$ as a constant leads, as expected, to commutators which cannot be expressed solely in terms of the generators of the integrity basis alone, but require the adjunction of (functionally) dependent polynomials. Proceeding along these lines, a long but routine computation shows that the minimal quadratic algebra that contains the (symmetrized) integrity basis (\ref{inta4}) is determined by the following nine polynomials of degree at most three:  

\begin{gather} 
A_{1}=\lambda E  + H ,\quad A_{2}= \frac{\lambda}{2}E^2  - 2 \frac{1}{\lambda}E F + E H,\quad A_{3} = 2\lambda E  + 4 E F + H^2 ,\quad A_{4}= P_{0}^2  - \frac{2}{\lambda}P_{1}P_{0}, \nonumber\\
A_{5}= -\frac{1}{2}m E - \frac{\lambda}{4}EP_{0}^2  + P_1 E P_0  - \frac{1}{\lambda} P_1^2  E ,\quad A_{6}= \frac{\lambda}{2}m E  + P_1 P_0- F P_0^2 + P_1 H P_0 + P_1^2  E,\nonumber\\
 A_{7}= -\frac{1}{2}\lambda^2 E^2  - \frac{\lambda^3}{4}E^3  + \lambda E^2 F - \frac{3}{4}\lambda^2 E^2 H + E H F - \frac{\lambda}{2}E H^2,\label{BSA}\\
A_8= 4\lambda E  + 3\lambda^2 E^2  + 12 E F + \lambda^3 E^3 +  3\lambda^2 E^2 H +  3\lambda  E H^2 + H^3,\quad A_9= \frac{\lambda}{2}E P_0^2  - \frac{2}{\lambda}F P_0^2+H P_0^2.\nonumber
\end{gather}

\noindent As $A_1=\mathcal{H}_0$ and $A_2$, $A_3$ are algebraically dependent, it is immediate to verify that $\left[ A_2,A_3\right]=0$ holds. For the remaining generators of the quadratic algebra, we obtain the commutators 

\begin{gather}
\left[A_2,A_4\right]= -\frac{2}{\lambda}\left(\left( 2 + A_1\right) A_4  +4 A_5  +2 \frac{1}{\lambda} A_6\right),\quad \left[A_2,A_6\right]=0, \quad [A_2,A_7]=0, \nonumber\\
\left[A_2,A_5\right]= \left(  \frac{2}{\lambda^2} + \frac{1}{2\lambda}A_1\right)m A_1 - \frac{1}{2\lambda^2}m A_3 + \frac{1}{\lambda}\left(4  + 3 A_1\right) A_4 + \frac{8}{\lambda}A_5, \nonumber\\
-\frac{1}{2\lambda} A_3 A_4 + \frac{1}{\lambda} A_1 A_5 - \frac{1}{\lambda^2} A_1 A_6 - \frac{4}{\lambda^2} A_6  + \frac{1}{2\lambda} A_1 A_9, \quad [A_2,A_8]=0   ,\nonumber\\
\left[A_2,A_9\right]=  \frac{4}{\lambda^2}m A_1(1 +  A_1)  - \frac{1}{\lambda^2}m A_3 + \frac{2}{\lambda}A_1 (A_4 -A_5)- \frac{1}{\lambda}A_3 A_4 - A_1\left( \frac{2}{\lambda^2}A_6 + \frac{1}{\lambda}A_9\right),\nonumber\\
\left[A_3,A_4\right]=8 A_4 + 4 A_1 A_4 + 16 A_5 +  \frac{8}{\lambda}A_6,\quad  [A_3,A_6]=0,\quad [A_3,A_7]=0,\quad [A_3,A_8]=0 \nonumber\\
\left[A_3,A_5\right]= - \frac{m}{\lambda} A_1(A_1+4 )  + \frac{m}{\lambda} A_3  + (A_3 - 6 A_1 -8)A_4  - 2( A_1 + 8)A_5  +  \frac{2}{\lambda}(A_1 - 4 ) A_6  - A_1 A_9, \nonumber\\
\left[A_3,A_9\right]= -\frac{8m}{\lambda} A_1  -  \frac{2}{\lambda}m A_1^2 +  \frac{2}{\lambda}m A_3  - 4 A_1 A_4 + 2 A_3 A_4 + 4 A_1 A_5 + \frac{4}{\lambda} A_1 A_6 + 2 A_1 A_9,\nonumber\\
\left[A_4,A_5\right]= \frac{m}{\lambda} A_4 -\frac{1}{2} A_4^2 +  \frac{4m}{\lambda} A_5 ,\quad [A_4,A_6]=-4 m A_4 - 2 m A_1 A_4 - 8 m A_5 -  \frac{4}{\lambda}m A_6, \nonumber\\
\left[A_4, A_7\right]= (4 +2 A_1) A_4 +(8  - 2 A_1) A_5 +  \frac{4}{\lambda} A_6 - A_1 A_9 ,\quad [A_4,A_8]=- 12 (A_1 +2) A_4  - 48 A_5 - \frac{24}{\lambda} A_6,\nonumber\\
\left[ A_4,A_9\right]=-\frac{1}{\lambda}m(6 +4 A_1) A_4 -A_4^2 -   \frac{8m}{\lambda} (A_5+ \frac{1}{\lambda} A_6), \quad \left[A_5,A_6\right]=
  \frac{m^2}{\lambda}( 2+ \frac{1}{2}A_1)A_1+\nonumber\\
   - \frac{m^2}{2\lambda} A_3 + 4 m A_4 +3 m A_1 A_4 - \frac{m}{2} A_3 A_4 + 8 m A_5 + m A_1 A_5 +  \frac{4}{\lambda}m A_6 - \frac{m}{\lambda} A_1 A_6 +\frac{m}{2} A_1 A_9,\nonumber
\end{gather}	

\begin{gather}
\left[A_5, A_7\right]= \frac{m}{\lambda}\left(-2  A_1 + \frac{5}{4} A_1 - \frac{1}{4} A_3 - \frac{1}{4}A_1 A_3\right)  + (A_3 -4  -4 A_1) A_4  + \left(\lambda A_2  + A_3  - 8\right) A_5  -   \nonumber\\
\left(\frac{4}{\lambda} + A_2\right) A_6 +\frac{m}{4\lambda} A_8 + (A_1  + \frac{1}{2\lambda} A_2) A_9,\nonumber \\
\left[A_5,A_8\right]=  \frac{3m}{\lambda} (A_1^2+4A_1  - A_3) + 3(8 + 6A_1 - A_3) A_4 +6( 8 + 6 A_1) A_5 +  \frac{6}{\lambda}(4 - A_1 )A_6 + 3 A_1 A_9 \nonumber\\
\left[A_5, A_9\right]=  \frac{m^2}{\lambda^2}(4A_1 +  A_1^2 - A_3) + \frac{m}{\lambda}\left( (6 + 5 A_1- A_3) A_4 + 2( 4 -  A_1) A_5\right)   + \frac{2m}{\lambda^2}(4 - 2 A_1) A_6 -\nonumber\\
 \frac{1}{\lambda} A_4 A_6 - A_4 A_5+\left( \frac{m}{\lambda} A_1 +\frac{1}{2} A_4 \right)A_9,\quad \left [A_6,A_7\right]=0,\quad [A_6,A_8]=0,\quad  [ A_7,A_8]=0, \nonumber\\
\left[ A_6,A_9\right]= \frac{m^2}{\lambda}(A_3 -4 A_1 - A_1^2)  + m (A_3 -2  A_1) A_4   + 2 m A_1 A_5 +  \frac{2}{\lambda} A_1 A_6 + m A_1 A_9\nonumber\\
\left[A_7,A_9\right]=  \frac{4m}{\lambda} A_1 + \frac{m}{2\lambda} (A_3- 5 A_1^2)  + \frac{1}{2\lambda}A_1 A_3 + (24  + 6 A_1 - 5 A_3 )A_4 + 2 (A_3 + 24 - 5A_1) A_5 + 6 \lambda\times\nonumber\\
\left( A_2 A_5  + 6A_6 - \frac{5}{3} A_1 A_6\right) + \left(6 A_2 +  \frac{2}{\lambda} A_3\right) A_6 + \left(A_4 - \frac{m}{2\lambda}\right) A_8 +\left(3 A_1  - \lambda A_2\right) A_9,\nonumber\\
\left [A_8,A_9\right]=  \frac{6m}{\lambda}\left( A_3 +4 A_1 -  A_1^2\right) + (6 A_3 - 2 A_1) A_4  + 6 A_1\left( A_5 + \frac{2}{\lambda}  A_6 +  A_9\right).\nonumber
\end{gather}\label{KBSA}
\noindent To a certain extent, it seems surprising that, in spite of the simplicity of the Hamiltonian, which as a linear combination of the generators of the Borel subalgebra, the resulting commutators in the quadratic algebra generated by the operators $A_i$ are so complicated. 
 
 \subsubsection{Extended Borel}

\noindent As the ``extended Borel" case we understand the basic Hamiltonian 
\begin{equation}
 \mathcal{H}_0=  H  +  E  +  E^2 +  H E  +  H^2,\label{EBS}
\end{equation}
that generalizes naturally the previous case to a genuinely quadratic element in the enveloping algebra.  In this case, the starting Hamiltonian $\mathcal{H}_0$ does not belong to the Schr\"odinger algebra $\widehat{S}(1)$. As observed earlier, we can still use the analytical approach to derive elements in the commutant. In this context, we observe that the condition for an element $Q$ to be in the commutant can be rewritten as follows 
\begin{equation}
\begin{split}
\left[\mathcal{H}_0,Q\right]= & \left[(E+H)^2-HE-E,Q\right]=(E+H)\left[E+H,Q\right]+\left[E+H,Q\right](E+H)\\
& -H\left[E,Q\right]-\left[H,Q\right]E- \left[E,Q\right]=0.\label{quad}
 \end{split}
\end{equation}
In particular, if $Q$ commutes with both $E$ and $H$, then it commutes with $\mathcal{H}_0$. There are one quadratic and one cubic polynomial that commute simultaneously with $E$ and $H$, given by\footnote{The non-symmetrized invariants are easily computed using the system obtained from (\ref{scsa}) for $a_1=1,a_2=0$ and $a_1=0,a_2=1$ respectively.}
\begin{equation}
\mathcal{C}_2=4EF+H^2-2H,\quad \mathcal{C}_3=EP_1^2-FP_0^2+P_1HP_0-\frac{1}{2}mH.\label{cabo}
\end{equation}
As $\mathcal{H}_0$ itself does not commute with either $E$ or $H$, the linear combination $\mathcal{C}_2-\mathcal{H}_0$ is a quadratic polynomial that belongs to the commutant of $\mathcal{H}_0$,  independent of the latter. In addition to $\mathcal{C}_3$, there is only one independent cubic polynomial that commutes with the Hamiltonian, given by $\mathcal{C}^{\prime}_3=\lambda(\left(4F  -E -H -1\right)P_0^2-P_1P_0)$ with $\lambda\in\mathbb{R}$. As this element does not commute with either $E$ or $H$, it is independent on $\mathcal{C}_3$. As the only fourth-order polynomial that simultaneously commutes with $E$ and $H$ is given by $\mathcal{C}_2^2$, we test whether the preceding quadratic and cubic elements of the commutant generate a quadratic algebra. In order to simplify the coefficients of the resulting commutators, we consider the following representatives for these operators: 
\begin{equation}
\begin{split}
M_1 = 3 E + H + E^2 + E H + H^2=\mathcal{H}_0,\quad &
M_2 = -\frac{3 E}{4} - \frac{3 H}{4} - \frac{E^2}{4} + E F - \frac{1}{4}E H, \nonumber\\
M_3 = \frac{\left(4F  -E -H -1\right)P_0^2}{4}-\frac{1}{2}P_1P_0,\quad & M_4 = -\frac{1}{2}H m + \frac{1}{2} P_1 P_0 - \frac{1}{4} \left(E + 
 H +1\right)P_0^2 + P_1 \left(H P_0 + P_1 E\right), \nonumber\\
 \end{split}
\end{equation}
The relations $\mathcal{C}_2=M_1+4M_2$ and $\mathcal{C}_3=M_4-M_3$ hold. It is further straightforward to verify that the commutators  $\left[\mathcal{C}_2,M_2\right]=\left[\mathcal{C}_3,M_2\right]=0$ are satisfied, a fact that allows us to simplify the computations. Now the commutator of $M_2$ and $M_3$ leads to an expression that cannot be written in terms of $M_1,\dots ,M_4$ alone, but requires the adjunction of the fourth-order polynomial
\begin{equation}
\begin{split}
 M_5 =& - \frac{3}{8} m H- \frac{1}{16} P_0^2 + \frac{7}{8} P_1 P_0 - \frac{5}{16} 
   E P_0^2  - \frac{3}{16} H P_0^2 - \frac{1}{2} P_1 E P_0  + \frac{3}{4} P_1^2 E- \frac{1}{8} E H P_0^2  + \frac{1}{2} H F P_0^2 - \frac{1}{8} H^2 P_0^2\nonumber\\
   &  -  \frac{1}{4} P_1 E^2 P_0 + P_1 E F P_0 - \frac{1}{4} P_1 E H P_0. 
 \end{split}  
 \end{equation}
With this new element, that does not commute with $E$ or $H$, the polynomials $\left\{M_1,\dots ,M_5\right\}$ generate a quadratic algebra with non-vanishing commutators 
\begin{gather}
\left[ M_2,M_3\right]=[M_2,M_4]= -m M_2+ \frac{3}{2}M_3 - 2 M_4 +2 M_5,\nonumber\\
\left[M_2,M_5\right]= \frac{1}{4} m M_2 + \frac{1}{8}m M_1+ \left( \frac{1}{2}M_1 +M_2 - \frac{9}{8}\right) M_3  +\left( \frac{3}{2} +M_2 \right)M_4 - \frac{3}{2} M_5,\nonumber \\
 \left[M_3,M_4\right]= 2 m^2 M_2 - 3 m M_3 + 4 m M_4 - 4 m M_5,\label{KEBS}\\
 \left[M_3,M_5\right]=\frac{7}{2} m^2 M_2 - \frac{15}{4} m M_3 - 2 m M_2 M_3 + \frac{11}{2} m M_4 - M_3 M_4 -5 m M_5,\nonumber\\
 \left[M_4,M_5\right]= 4 m^2 M_2+ \frac{1}{4}m^2 M_1 - 6 m M_3 + m M_1 M_3 + \frac{17}{2} m M_4 + 2 m M_2 M_4 - M_3 M_4 - 8 m M_5.\nonumber
\end{gather}
It follows at once from these relations that the quadratic algebra is minimal, as no subset of $\left\{M_1,\dots ,M_5\right\}$ generates a non-Abelian algebra. As in the previous cases, a general Hamiltonian $H= H_0 + {\bf C}_0 $ that preserves the structure of the quadratic algebra can be defined, with explicit realization determined by the differential operators (\ref{REA1}). 

\subsubsection{Extended Cartan solvable}

\noindent Just as the elements $H,E$ generate the Borel subalgebra of $\mathfrak{sl}(2,\mathbb{R})$, the pair $H,P_1$ generates also a non-Abelian solvable two dimensional Lie algebra. In spite of this analogy, the generators behave rather differently seen as elements of the Schr\"odinger algebra $\widehat{S}(1)$, and thus it is expected that a minimal quadratic algebra obtained as the commutant of a Hamiltonian based on $H$ and $P_1$, if existing, will exhibit quite different properties to that previously obtained. With some abuse of notation, we will call this case, where the basic Hamiltonian is taken as 
\begin{equation}
\mathcal{H}_0= H  +  P_1  +  P_1  H  +  H^2 +  P_1^2, \label{ECSS} 
\end{equation}
the extended Cartan solvable case. For the system of PDEs formed by equation (\ref{inca}) and 
\begin{equation}
\widehat{P_1}(\varphi)= p_1 \frac{\partial \varphi}{\partial h}-p_0 \frac{\partial \varphi}{\partial e}+m\frac{\partial \varphi}{\partial p_0}=0, \label{inca2}
\end{equation} 
any integrity basis is formed by four elements, which can be chosen as the polynomials
\begin{equation}
Q_1=p_0p_1-mh,\quad Q_2=fp_0^2+2mef,\quad Q_3=\frac{1}{2}mh^2-hp_0p_1-ep_1^2,\quad Q_4=m.\label{inta42}
\end{equation}
Considering their symmetrization and adding the starting Hamiltonian (\ref{ECSS}), we analyze whether the polynomials 
\begin{equation}
\begin{split}
M_1 =& H  +  P_1  +  P_1  h  +  H^2 +  P_1^2=\mathcal{H}_0, \quad
M_2 =  P_1 P_0-m H , \nonumber\\
M_3 =&  \frac{m}{2} P_1  + \frac{m}{2} P_1H + \frac{m}{2} P_1^2 + P_1 H P_0 + P_1^2 E, \quad M_4 = -2 m H + 2 m E F + F P_0 ^2, 
 \end{split}
\end{equation}
close as a quadratic algebra. The commutators are given by 
\begin{equation}
\begin{array}[c]{ll}
\left[M_1,M_2\right]=[M_1,M_3]=[M_1,M_4]=0,& [M_2,M_3]=m M_2-M_2^2+m^2 M_1-2 m M_3,\\
\left[M_2,M_4\right]= m M_2-M_2^2+m^2 M_1-2 m M_3,& [M_3,M_4]= -m M_2+M_2^2-m^2 M_1+2 m M_3.
\end{array}
\end{equation} 
We observe that $\left\{M_1,M_2 ,M_3\right\}$ are sufficient to generate a minimal quadratic algebra, as the commutators do not depend on
$M_4$. The adjunction of the latter polynomial provides a higher dimensional quadratic algebra, but the commutators all have the same structure. This shows that a minimal quadratic algebra does not necessarily contain the whole integrity basis. If we now add a fifth element $M_5=m M_2-M_2^2+m^2 M_1-2 m M_3$, that corresponds to a fourth-order polynomial in the generators, the relation $\left[M_4-M_3-M_2,M_5\right]=0$ implies that  
\begin{gather}
\left[M_2,M_5\right]=-2m M_5,\quad [M_3,M_5]=-m M_5+2M_2M_5,\quad
\left[M_4,M_5\right]= m M_5+2M_2M_5.
\end{gather} 
It follows at once from these commutators that the polynomials $\left\{M_1,M_3-M_4,M_5\right\}$ do no more generate a quadratic algebra, but a solvable Lie algebra isomorphic to $\mathfrak{b}\oplus\mathbb{R}$.\footnote{For general properties of decomposable Lie algebras, see \cite{Sn14}.}  

\section{The Conformal Galilean algebra  $\widehat{S}(3)$ }

Like $\widehat{S}(1)$, the Lie algebra $\widehat{S}(3)$ corresponding to the value $j=\frac{3}{2}$, is a semidirect sum of the simple Lie algebra $\mathfrak{sl}(2,\mathbb{R})$ and a Heisenberg  algebra $\mathfrak{h}_2$,  where in this case $\dim\mathfrak{h}_2=5$ and the characteristic representation is given by $D_{\frac{3}{2}}\oplus D_0$. Over the basis $\left\{H,E,F,P_0,P_1,P_2,P_3,m\right\}$ the brackets are given by 
\begin{equation}\label{CGA}
\begin{array}[c]{llllll}
\left[ H,E\right]=2 E, & [H,F]=-2 F,&  [E,F]=H,&  [H,P_{0}]=3P_0,& [H,P_{1}]=  P_{1},& [H,P_{2}]=-P_{2}, \\
\left [H,P_{3}\right]=-3 P_3,& [E,P_0]=0,& [E,P_1]=P_0,& [E,P_2]=2P_1,& [E,P_3]=3P_2,& [F,P_0]=3P_1, \\
\left [F,P_{1}\right]=2P_2, & [F,P_2]]=P_3,& [F,P_3]=0,& [P_0,P_3]=6m,& [P_1,P_2]=-2m. &
\end{array}
\end{equation}
$\widehat{S}(3)$ admits two Casimir operators \cite{C40}, with a non-central invariant of order four and explicit expression
\begin{equation}\label{cas4}
\begin{split}
{\bf C}_0= & P_1^2P_2^2-\frac{1}{3}P_0^2P_3^2-\frac{4}{3}\left(P_1^3P_3+P_0P_2^3\right)+2P_0P_1P_2P_3+\frac{4m}{3}H(P_1P_2-P_0P_3)
+\frac{8m}{3}E(P_2^2-P_1P_3)\\
 & +\frac{8m}{3}F(P_0P_2-P_1^2)-2m(P_1P_2-3P_0P_3)-\frac{4m}{3}(H^2+4EF)+8m^2H.
\end{split}
\end{equation}
We shall see that, among the choices of algebraic Hamiltonians considered for $\widehat{S}(1)$, not all generalize as expected to $\widehat{S}(3)$, due to the structure of the $\mathfrak{sl}(2,\mathbb{R})$-representation describing the semidirect sum. 

\subsection{Cartan case}
As $\mathcal{H}_0=H$ belongs to the Lie algebra, we can determine the commutant via the coadjoint representation and the symmetrization map. For the differential operator $\widehat{H}$ associated to $H$, the PDE
\begin{equation}
\widehat{H}(\varphi)= 2e \frac{\partial \varphi}{\partial e}-2f \frac{\partial \varphi}{\partial f}+3p_0\frac{\partial \varphi}{\partial p_0}+p_1\frac{\partial \varphi}{\partial p_1}-p_2\frac{\partial \varphi}{\partial p_2}-3p_3\frac{\partial \varphi}{\partial p_3}=0 \label{inca32}
\end{equation} 
admits an integrity basis formed by the seven elements
\begin{gather}
Q_1=h,\quad Q_2=ef,\quad Q_3=p_0p_3,\quad Q_4=p_1p_2,\quad Q_5=ep_1p_3,\quad Q_6=fp_1^2,\quad Q_7=m,\nonumber
\end{gather}
and thus the commutant in the enveloping algebra is generated by the symmetrization of these elements, which can be taken after simplification (and neglecting $m$) as 
\begin{gather}
A_1=H,\quad A_2=EF,\quad A_3=P_3P_0,\quad A_4=P_2P_1,\quad A_5=P_3EP_1,\quad A_6=FP_1^2.\label{inca33}
\end{gather}
However, in contrast to the case  $\widehat{S}(1)$, for $\widehat{S}(3)$ the polynomials $A_1,\dots ,A_7$ in the commutant of the Cartan generator $H$ do not lead to a quadratic algebra, but to a cubic one. This can be verified as follows: the commutator of the elements $A_2$ and $A_4$ contains the monomial $Z_1=EP_2^2$, which cannot be expressed as a product of two elements in (\ref{inca33}), and should therefore be adjoined to the generators. Now  $\left[Z_1,A_6\right]$ shows that the polynomial $Z_2=P_2^2P_1^2H$ must also be added to the generators if the algebra is to be quadratic. On the other hand, the commutator $\left[Z_1,Z_2\right]$ implies that $Z_3=P_2^4P_1P_0H$ must also belong to the set of generators. Evaluating recursively the commutators of $Z_1$ and $A_6$ with $Z_k$ for $k\geq 2$, we find that the monomials of the type $P_2^{\alpha+3\beta}P_1^{\alpha}P_0^{\beta}H$ with $\alpha,\beta\in \mathbb{N}$ must be taken as generators, hence leading to a  quadratic algebra that is not finitely generated. 
The reason for the failure of the construction of a finite dimensional quadratic algebra is ultimately a consequence of the representation theory of the simple Lie algebra $\mathfrak{sl}(2,\mathbb{R})$. For $\widehat{S}(1)$ the only elements  in the commutant of $H$ that mix one generator of the simple part and generators of the radical are of the type $P_1^{a}P_0^aH^b$, $E^{a}P_1^{2a}$ and $F^{a}P_0^{2a}$. In particular, the third-order monomials of this type are $P_1P_0H$, $EP_1^{2}$ and $FP_0^{2}$, the commutator of which can be easily expressed as a quadratic product. In addition, the only elements in $C_{\mathcal{U}(\mathfrak{g}}(H)$ of even order and depending on the generators $P_0,P_1$ are powers of $P_0P_1$, while for any representation $D_{\frac{j}{2}}$ with $j\geq 3$ there are $j$ generators $P_sP_{j+1-s}$ of degree $2$, so that the number of even-order generators increases exponentially. 

\medskip
\noindent Albeit there does not exist a minimal quadratic algebra associated to $\widehat{S}(3)$ that contains the generators of the integrity basis (\ref{inca33}), we can always find a quadratic subalgebra that only involves a subset of these generators. For instance, the key observation to find such algebras is to exclude those generators that contain $F$, in order to avoid the generators of the type $Z_k$ above. One possible quadratic algebra generated by six elements is given by 
\begin{gather}
B_1=H,\quad B_2=P_3P_0,\quad B_3=P_2P_1,\quad B_4=P_3EP_1,\quad B_5=P_3P_1^3,\quad B_6=P_2^2P_1^2,\label{inca36}
\end{gather}
where the non-vanishing commutators are
\begin{gather}
\left[B_2,B_4\right]=6m B_4 -3B_2B_3,\quad [B_2,B_5]=6m B_5,\quad
\left[B_3,B_4\right]= 2(m B_4-B_5)-B_2B_3,\nonumber\\
 \left[B_3,B_5\right]=6m B_5,\quad \left[B_4,B_5\right]=3(B_2B_5+B_3B_5)-18mB_5,\nonumber\\
\left[B_4,B_6\right]=2B_2B_6-4m(B_2B_3+B_3B_4)+4B_3B_5-12mB_5,\quad 
\left[B_5,B_6\right]=24m^2B_5-12mB_3B_5.\nonumber
\end{gather} 

\medskip
\noindent In any case, this shows that for the Cartan case there is no obvious choice of generators for a quadratic algebra, as the latter
has to be chosen as a suitable quadratic subalgebra of the cubic algebra generated by the polynomials of the integrity basis (\ref{inca33}). 
For this reason, we skip the analysis of the Borel case, as it gives rise, similarly, to an ambiguity concerning the choice of generators. 

\subsection{Extended Borel}

In order to analyze the existence of a finitely generated quadratic algebra contained in the commutant of the Borel polynomial (\ref{EBS}), we proceed again using equation (\ref{quad}), trying to obtain polynomials related to the operators that commute with the generators intervening in the expression of $\mathcal{H}_0$. Using the coadjoint representation, we can easily find a quadratic, a cubic and a fourth-order polynomial that commute simultaneously with $E$ and $H$. After symmetrization, these are given by 
\begin{equation}\label{cabo2}
\begin{split}
\mathcal{C}_2=4EF+H^2-2H,\quad \mathcal{C}_3=\frac{H}{2}\left(P_1P_2-P_0P_3\right)+E(P_2^2-P_1P_3)+F(P_0P_2-P_1^2)+2mH,\\
\mathcal{C}_4=P_1^2P_2^2-\frac{4}{3}\left(P_0P_2^3+P_1^3P_3\right)+2P_0P_1P_2P_3-\frac{1}{3}P_0^2P_3^2-2m\left(P_1P_2-3P_0P_3\right). 
 \end{split}
\end{equation} 
Now, as $\left[\mathcal{H}_0,H\right]\neq 0$, the operator $\mathcal{C}_2$ is independent of the Hamiltonian. In analogy to the case for $\widehat{S}(1)$, we make the following choice for the quadratic generators in the commutant: 
\begin{equation}
M_1 = 3 E + H + E^2 + E H + H^2=\mathcal{H}_0, \quad
M_2 = -\frac{1}{4}\left(3E +3H +E^2+ EH\right) + E F. \nonumber\\
\end{equation}
A third-order polynomial in the commutant $C_{\mathcal{U}(\mathfrak{g}}(\mathcal{H}_0)$ that is independent of $\mathcal{C}_3$ can be found restricting to operators that are at most linear in the generators $H,E,F$. One possible choice is given by 
\begin{equation}
M_3 =\frac{1}{4} \left(4F  -H- E   -1 \right)P_1^2 +\frac{1}{2}P_0P_1-\frac{1}{4}P_2P_1-\frac{3}{4}P_3P_0+ \frac{1}{4} \left(P_2 H +P_2E- 4P_2 F\right)P_0.\nonumber
\end{equation}
If we now compute the commutator $[M_2,M_3]$, we observe that the resulting fourth-order operator cannot be written in terms of $M_1,M_2,M_3$ and $\mathcal{C}_4$, implying the existence of another element in the commutant having degree four and not commuting with either $H$ or $E$. Iteration of the latter commutator shows similarly that there is a second third-order polynomial independent of $\mathcal{C}_3$ that commutes with $\mathcal{H}_0$. These additional operators can be taken as follows:  
\begin{equation}
{\small 
\begin{split}
 M_4 =&  \frac{7P_3P_0+P_2P_1-P_1^2+2P_0P_1}{4}  -\frac{(E+H)P_1^2}{4}+\frac{P_2(E+H)P_0}{4}+
 \frac{P_2HP_1-P_3HP_0}{2} +P_2^2 E-P_3EP_1-2 mH , \nonumber\\
M_5 =&  \mathcal{C}_4, \quad
M_6=  \frac{1}{2} \left(P_2 -P_0  + \frac{1}{2} (2HP_0 -2P_2E) + 
  2P_3 E + 2P_2 EF- \frac{1}{2} P_2 (E^2 + EH)-\frac{3}{2} P_2 H \right)P_1+\frac{1}{4}P_1^2+2P_3P_0+  \nonumber\\
  & \frac{\left(4 HF - H^2   - E  - EH\right) P_1^2}{4}    +\frac{\left( 2 P_2 E   -  (P_2 + P_3) H\right)P_0}{4}   + 
  \frac{P_2\left( EH   - 4HF  + H^2 \right)P_0}{4} +   \frac {P_3 \left(E - 4EF + EH\right) P_0}{4}.
 \end{split} }
\end{equation}
These six polynomials are sufficient to  define a quadratic algebra, with non-vanishing commutators
\begin{gather}
\left[M_2,M_3\right]=[M_2,M_4]=-\frac{1}{2} [M_3,M_4]=-4 m M_2+\frac{3}{2} M_3-\frac{1}{2}M_4+M_6,\nonumber\\
\left[M_2,M_6\right]= (8 M_2+M_1)-\frac{9}{2} M_3+(2 M_2 +M_1) M_3+\frac{3}{2} M_4-3 M_6+2 M_2 M_4,\label{KEBS2}\\
\left[M_3,M_6\right]=(16 m M_2-3 M_3-4  M_2 M_3+2 M_4-5  M_6)m+\frac{9}{8} M_5-2 M_3 M_4+3 M_2 M_5,\nonumber\\ 
\left[M_4,M_6\right]= 2 m^2 (M_1+16 M_2)+(5  M_4-12 M_3-11 M_6+2  M_1 M_3 +4 M_2 M_4)m+\frac{9}{8} M_5+3 M_1 M_5-2 M_3 M_4. \nonumber
\end{gather}
 The algebra is minimal in the sense that we cannot skip one of the generators.

\subsection{Extended Cartan solvable}

\noindent As  for $\widehat{S}(3)$ the $\mathfrak{sl}(2,\mathbb{R})$-representation $D_{\frac{3}{2}}$ is four dimensional, we take 
\begin{equation}
\mathcal{H}_0 =   h  +   P_3   +  P_3^2    +    h P_3   +   h^2  +  P_2  + 
  h P_2  +  P_2^2  + P_2 P_3.\label{ECS2}
\end{equation}
as starting Hamiltonian, i.e., the generator of the Cartan subalgebra and the elements in the representation having negative weight with respect to $H$.   This choice ensures that the number of commuting polynomials is not too high, hence allowing to identify a low-dimensional quadratic algebra. As before, we first consider the polynomials that commute simultaneously with $H$, $P_2$ and $P_3$. Using the analytical approach, it can be shown that, up to order five, there are four independent solutions to the system, corresponding to one quadratic and three fourth order polynomials, that after symmetrization can be chosen as:
\begin{equation}
\begin{split}
Q_1 = & P_3 P_0- P_2 P_1 +2 m H  ,\\
Q_2 = &P_0P_1P_2P_3-\frac{1}{2}\left(P_0^2P_3^2+P_0P_2^3\right)+m(EP_2^2+2H(P_1P_2-P_0P_3))+ 2mP_0P_3+2m^2H^2-m^2H,\nonumber\\
Q_3= &   P_0P_1P_2P_3-P_1^3P_3-2m(EP_1P_3+FP_0P_2-FP_1^2)+2mP_0P_3-4m^2EF+4m^2H,\\
Q_4= & 2 m P_2^2 E + P_2^2 P_1^2 - P_2^3 P_0.
\end{split}
\end{equation}
In particular, no cubic or fifth order polynomial commutes simultaneously with $H$, $P_2$ and $P_3$. Analyzing the commutators of the five polynomials $M_1=\mathcal{H}_0$, $M_2=Q_1$, $M_3=Q_2$, $M_4=Q_3$, $M_5=Q_4$ in the commutant of $\mathcal{H}_0$, the following relations for $2\leq i\leq5$ are obtained:  
 \begin{equation}
\begin{array}[c]{l}
\left[M_2,M_3\right]=-mM_5,\quad [M_2,M_4]=2mM_5,\quad [M_2,M_5]=4mM_5, \\
\left[M_3,M_4\right]= -24m^2 M_5-3mM_2M_5,\quad [M_3,M_5]= 8m^2 M_5-2m M_2M_5,\\
\left[M_4,M_5\right]= -28m^2 M_5-4mM_2M_5.
\end{array}
\end{equation}  
A minimal quadratic algebra is given by the polynomials $M_1,M_2,M_4,M_5$. If to these we add the (algebraically dependent) polynomial $M_6=M_2M_5$, then we get another structurally more interesting quadratic algebra, with new commutators 
\begin{equation}
\begin{array}[c]{l}
\left[M_2,M_6\right]=4mM_6,  \\
\left[M_4,M_6\right]= -160m^3 M_5-64m^2M_6-4mM_4M_5-6M_2M_6,\\
\left[M_4,M_5\right]= -320m^3 M_5-72m^2M_6-8mM_4M_5-4M_2M_6.
\end{array}
\end{equation}  
In analogy to the case $\widehat{S}(1)$, the polynomials $\left\{M_1,M_2,M_6\right\}$ generate a solvable Lie algebra isomorphic to $\mathfrak{b}\oplus\mathbb{R}$.  
 
 \medskip
 \noindent As in the case of the (1+1)-Schr\"odinger algebra, an explicit realization by differential operators can be obtained and used to define a Hamiltonian 
$\mathcal{H}=\mathcal{H}_0+{\bf C}_0$, from which explicit expressions for the integrals (i.e., the polynomials in the commutant generating a quadratic algebra) can be immediately deduced.  
For $\widehat{S}(3)$, a realization as vector fields in $\mathbb{R}^3$ is given by (see \cite{Ai13}):  
\begin{equation}\label{REA3}
\begin{array}[c]{ll}
H=-2 t \partial_t -3 x_0 \partial_{x_{0}} - x_1 -\frac{1}{2},&  F= -t^2 \partial_t -3 x_0 \partial_{x_{0}} -t x_1 \partial_{x_{1}} -\frac{1}{2} t + 2 m x_1^2 -3 x_{0} \partial_{x_{1}}\\
E=\partial_t,\quad P_0=- \partial_{x_{0}}, & P_1=-t \partial_{x_{0}}- \partial_{x_{1}},\quad P_2= 2 m x_1 - t^2 \partial_{x_{0}}-2 t \partial_{x_{1}},\\
P_{3}=6 m (t x_{1}-x_{0})  - t^3 \partial_{x_{0}}- 3 t^2 \partial_{x_{1}}. &
\end{array}
\end{equation}  

\section{Extended Cartan solvable case for $\widehat{S}(n)$}

The comparison of the results obtained for $\widehat{S}(1)$ and $\widehat{S}(3)$ opens the question whether a (minimal) quadratic algebra can be identified for the Schr\"odinger algebra $\widehat{S}(n)$ for arbitrary values $n\geq 5$. We recall that the semidirect sum structure of $\widehat{S}(n)$ is given by $
\mathfrak{sl}(2)\overrightarrow{\oplus}_{D_{\frac{j}{2}}\oplus D_0}\mathfrak{h}_{\frac{2j+1}{2}}$, where $n=2j$, with commutators  
\begin{equation}\label{CGAN}
\begin{array}[c]{lll}
\left[ H,E\right]=2 E, & [H,F]=-2 F,&   \left[H,P_{s}\right]=(2j-2s)P_s,\; ( 0\leq s\leq 2j),  \\
\left[E,F\right]=H, & \left [E,P_{s}\right]= s P_{s-1}, & 1\leq s\leq 2j, \\
\left [F,P_s\right]=(2j-s)P_{s+1},& 0\leq s\leq 2j-1,&    \\
\left [P_r, P_s\right]=\delta_{r+s,2j} I_r m, & I_r =(-1)^{r+j+\frac{1}{2}} (2j-r)! r!, & 0\leq r<s\leq 2j
\end{array}
\end{equation}
over the basis $\left\{H,E,F,P_0,\dots, P_{2j},m\right\}$. Although the answer is in the affirmative, the preceding analysis shows that the choice of generators is by no means obvious, at least for the Cartan and the Borel cases. The cases of the extended Cartan and Borel polynomials are more promising, although the computational difficulties increase considerably. 

\medskip
\noindent The extended Cartan case, however, constitutes a sort of special case, as we can infer the existence of a minimal quadratic algebra without explicitly making computations in the enveloping algebra $\mathcal{U}(\widehat{S}(n))$, but arguing with the analytical approach. The suitable algebraic Hamiltonian $\mathcal{H}_0$ to be considered in this case is 
\begin{equation}\label{genha}
\mathcal{H}_0=H+ H^2+\sum_{\ell=\frac{2j+1}{2}}^{2j} \left(P_\ell + P_\ell^2 +HP_\ell\right)+\sum_{\frac{2j+1}{2}\leq \ell <\ell^{\prime}\leq 2j} P_\ell
P_\ell^{\prime}
\end{equation}
We have seen for the values $j=\frac{1}{2},\frac{3}{2}$ that a quadratic algebra can be determined considering those elements in  $C_{\mathcal{U}(\mathfrak{g}}(\mathcal{H}_0)$ that commute simultaneously with $H$ and the generators $P_s$ having a negative eigenvalue with respect to the Cartan subalgebra. We proceed along these lines for general values of $j$. In terms of the basis (\ref{CGAN}), the operators commuting with $H$ and $P_s$ for $s\geq \frac{2j+1}{2}$ are obtained symmetrizing the polynomial solutions of the system ($\frac{2j+1}{2}\leq \ell\leq 2j$):
\begin{equation}\label{snsy}
\begin{split}
\widehat{H}(\varphi) = &2e \frac{\partial \varphi}{\partial e}-2f \frac{\partial \varphi}{\partial f}+\sum_{s=0}^{2j} (2j-2s)p_s \frac{\partial \varphi}{\partial p_s}=0,\\
\widehat{P}_\ell(\varphi) = & 2(\ell-j)p_\ell \frac{\partial \varphi}{\partial h}-\ell p_{\ell-1}\frac{\partial \varphi}{\partial e}-(2j-\ell)p_{\ell+1}\frac{\partial \varphi}{\partial f} +I_{\ell}m \frac{\partial \varphi}{\partial p_{2j-\ell}}=0.
\end{split}
\end{equation}
For each $j=2j_0-1\geq 1$, the system (\ref{snsy}) has $j_0+1$ independent solutions. Let $[\alpha]=\left[\alpha_0,..,\alpha_{2j+4}\right]\in\mathbb{N}^{2j+4}$ and suppose that the homogeneous polynomial
\begin{equation}
P\left(h,e,f,p_0\dots ,p_{2j}\right)=\lambda_{[\alpha]}p_0^{\alpha_0}\dots p_{2j}^{\alpha_{2j}}h^{\alpha_{2j+1}}e^{\alpha_{2j+2}}f^{\alpha_{2j+3}}m^{\alpha_{2j+4}}
\end{equation}
is a solution of the system. As $H$ acts diagonally on each generator of the Lie algebra, the polynomial $P$ must satisfy the numerical constraint 
\begin{equation}
\sum_{s=0}^{2j} (2j-2s)\alpha_{s}+2\alpha_{2j+2}-2\alpha_{2j+3}=0.\label{eiw}
\end{equation}
As the Hamiltonian $\mathcal{H}_0$ does not satisfy this condition, it is not a solution of the system. An integrity basis for the solutions of  the PDE $\widehat{H}(\varphi) =0$ is given by the polynomials
\begin{equation}
\left\{h,\; m,\; ef,\; p_sp_{2j-s}\; (0\leq s\leq [j]),\; ep_{\frac{2j+1}{2}}^2,\; p_rp_{\frac{2j+1}{2}}^{2j-2r}\; (0\leq r\leq \frac{2j-3}{2})\right\}.
\end{equation}
A routine but cumbersome computation shows that the system (\ref{snsy}) admits, besides the central element $m$, only one quadratic solution
\begin{equation}
J_2=\sum_{s=0}^{\frac{2j-1}{2}}\alpha_s (-1)^s p_sp_{2j-s}-(-1)^{\frac{2j-1}{2}}mh,\label{sekca}
\end{equation}
where $\alpha_0=1$ and $\alpha_s=\Omega\left({\frac{2j+3}{2}}-s,s+1\right)$ for values $s\geq 1$, with $\Omega$ denoting the generating function 
\begin{equation}
\Omega\left(\ell,q\right)= \frac{4^{q-1}\; \Gamma\left(\frac{2\ell-1}{2}+q-1\right)\;\Gamma(2\ell)}{\Gamma(\ell)\; \Gamma\left(\frac{2\ell-1}{2}\right)\;\Gamma(2\ell+q-1) }\prod_{s=0}^{\ell-2}(s+q),\label{genfunc}
\end{equation}
where $\Gamma(z)$ is the usual Gamma function \cite{Abs}. Now, a cubic polynomial that satisfies the first equation of the system must have the form 
\begin{equation}
J_3=\sum_{s+u=2j}\left(\beta_s^1e p_sp_{u+1} +\beta_s^2f p_sp_{u-1} +(\beta_s^3h+\beta_s^4m)p_sp_{u}\right)+\sum_{s=0}^{3}\beta_s^5 h^{3-s}m^s
+(\beta_1^6h+\beta_2^6m)ef
\end{equation}
due to the numerical constraint (\ref{eiw}). Inserting this expression into the remaining $\frac{2j+1}{2}\leq \ell\leq 2j$ equations and evaluating the resulting coefficient system implies the following relations
\begin{equation}
\beta_s^1=0,\quad \beta_s^2=0,\quad \beta_s^3=0,\quad \beta_s^4=(-1)^s\alpha_s\mu,\quad \beta_5^0=\beta_5^1=0,\quad \beta_5^2=(-1)^{\frac{2j+1}{2}}\mu,\quad\beta_6^1=\beta_6^2=0,\quad \mu\in\mathbb{R}.
\end{equation}
Therefore $J_3=\mu mJ_2+\beta_3^5m^3$ and no independent cubic solution exists. A similar argumentation holds for higher odd orders, i.e., it follows that any polynomial $P$ of odd order that is a solution of (\ref{snsy}) is a product of $m$ and some even-order polynomials. If we consider $m$ as a constant, then no odd-order solution to the system exists. This means that any maximal independent set of polynomials (or an integrity basis) is formed by polynomials of even degree. That the system possesses at least two independent fourth-order solutions for $n\geq 5$ can be shown observing that the equations $\widehat{P}_\ell(\varphi) =0$ admit the three solutions\footnote{The values $j=\frac{1}{2}$ and $\frac{3}{2}$ are special cases, as the first admits cubic polynomials and the second three fourth-order solutions.} 
\begin{equation}
\begin{split}
I_{1}=&p_{\frac{2j+1}{2}},\quad I_{2}=\sum_{s=0}^{\frac{2j-3}{2}} (-1)^s\left(\begin{array}[c]{c} 2j-1\\s\end{array}\right)p_sp_{2j-1-s}+\frac{(-1)^{\frac{2j-1}{2}}}{2}\left(\begin{array}[c]{c} 2j-1\\\frac{2j-1}{2}\end{array}\right)p_{\frac{2j-1}{2}}^2+\frac{(-1)^{\frac{2j-1}{2}}}{2}(2j-1)!em,\nonumber\\
I_{3}=&\sum_{s=1}^{\frac{2j-1}{2}} (-1)^s\left(\begin{array}[c]{c} 2j-1\\s-1\end{array}\right)p_sp_{2j-s}+\frac{(-1)^{\frac{2j-1}{2}}}{2}\left(\begin{array}[c]{c} 2j-1\\\frac{2j-1}{2}\end{array}\right)p_{\frac{2j-1}{2}}^2+\frac{(-1)^{\frac{2j-1}{2}}}{2}(2j-1)!fm.\nonumber
\end{split}
\end{equation}
These functions do not satisfy the first equation of the system, but following relations hold:
\begin{equation}
\widehat{H}(I_1)=-2 I_1,\quad \widehat{H}(I_2)=2 I_2,\quad \widehat{H}(I_3)=-2 I_3,\nonumber 
\end{equation}
hence guaranteeing that $J_{41}=I_1^2I_2$ and $J_{42}=I_2I_3$ are always solutions of (\ref{snsy}). There is a third fourth-order solution $I_{43}$ that does not arise from the polynomials commuting solely with the $P_s$, in analogy to $J_2$, the explicit expression of which is skipped because of its length.\footnote{The coefficients of this solution, not yet determined generically, are also described in terms of generating functions of the type (\ref{genfunc}).} Let $A_2=\Phi(J_2)$, $A_3=\Phi(I_{41})$, $A_4=\Phi(I_{42})$ and $A_5=\Phi(I_{43})$ be the corresponding symmetrizations of $J_2$, $I_{41}$, $_{42}$ and $I_{43}$ respectively. It is not difficult to verify that 
\begin{equation}
\left[A_r,A_s\right]\neq 0,\quad  2\leq r<s\leq 5\label{gle}
\end{equation}
hold. Now, using the fact that there do not exist solutions of odd order (in the sense commented above), the preceding commutators must be elements of order four in the commutant of $\mathcal {H}_0$ (see equation (\ref{grad})). As $A_2^2$, $A_3$, $A_4$ and $A_5$ exhaust the polynomials of fourth order commuting with $\mathcal{H}_0$, the commutators in (\ref{gle}) involving $A_2$ must be combinations of them. It is computationally harder to verify that  the commutators $\left[A_{3},A_{4}\right]$, $\left[A_{3},A_{5}\right]$ and $\left[A_{4},A_{5}\right]$ can also be written in terms of $A_2,A_3,A_4,A_5$, and that these commutators are genuinely quadratic.\footnote{These commutators have been determined and verified using a symbolic computer package for the values $2j=5,7,9,11$ and $13$. The obtainment of an explicit formula for all values of $j$ is still an unsolved problem.} We conclude that a minimal quadratical algebra is determined by the generators of the set $\left\{\mathcal{H}_0,A_2,A_3,A_4,A_5\right\}$. It may be observed that adjoining higher-order polynomials (of even degree), higher-dimensional quadratic algebras may be constructed. 

 \medskip
 \noindent 
Using the generic realization depending on a free parameter $\delta$ given by (see e.g. \cite{Ai13}) 
\begin{equation}
\begin{split}
H&=\delta -2t \partial_t  -\sum_{j=0}^{l-\frac{1}{2}} 2(l-j) x_j \partial_j,\quad F=  -t (H) + t^2 \partial_t  + \frac{m}{2}((l+\frac{1}{2})!)^2 x_{l-\frac{1}{2}}^2  - \sum_{j=0}^{l-\frac{1}{2}} (2l -j) x_j \partial_{j+1},\\
E&= \partial_t,\quad P_i=-\sum_{\ell=0}^{k} \begin{pmatrix} k \\ j \end{pmatrix} t^{k-j} \partial_{x_j},\quad  i=0,...,l-\frac{1}{2},\\
P_s&= M \sum_{\ell=2j-k}^{l-\frac{1}{2}} \begin{pmatrix} k \\ 2l-j \end{pmatrix} I_{2l-j} t^{k-2l+j} x_j - \sum_{j=0}^{l-\frac{1}{2}} \begin{pmatrix} k \\ j \end{pmatrix} t^{k-j} \partial_{x_j},\quad  i=l+\frac{1}{2},...,2l
\end{split}
\end{equation}
explicit examples for these algebraic Hamiltonian and  the corresponding (noncommuting) integrals can be obtained.

\section{Conclusions}

\noindent In this work, we have considered the construction of quadratic algebras associated to an algebraic Hamiltonian $\mathcal{H}_0$ defined in terms of the generators of a Lie algebra analyzing the commutant  $C_{\mathcal{U}(\mathfrak{g}}(\mathcal{H}_0)$ of $\mathcal{H}_0$ in the enveloping algebra $\mathcal{U}(\mathfrak{g})$. This construction, albeit it differs substantially from that considered in \cite{cam21}, in the sense that it is not based at all on realizations of Lie algebras by vector fields, still uses some elements of the analytical approach, as suitable generators of quadratic algebras contained in $C_{\mathcal{U}(\mathfrak{g}}(\mathcal{H}_0)$ are chosen in connection with polynomials commuting with those generators of $\mathfrak{g}$ intervening in the expression of the Hamiltonian $\mathcal{H}_0$. This allows, at least in principle, to start from a suitable set of polynomials contained in the integrity basis of such systems, to which eventually additional (dependent) polynomials are added to obtain a set of polynomials, the commutators of which close quadratically.  The procedure has been tested with success for some of the Schr\"odinger algebras $\widehat{S}(n)$, where various choices of algebraic Hamiltonian have been considered. It turns out that the representation of  $\mathfrak{sl}(2,\mathbb{R})$ describing the semidirect sum structure of $\widehat{S}(n)$ imposes some restrictions on some of these Hamiltonians, in the sense that, although it does not exclude the possibility of quadratic algebras to exist, the choice of suitable generators in the commutant is by no means obvious. Other cases, on the contrary, show some general pattern that can be systematized, as has been shown for the case of the extended Cartan case. A problem still to be solved in connection with this case is an explicit and generic description of the minimal quadratic algebra based on the polynomials commuting simultaneously with the Cartan subalgebra $H$ and the generators $P_s$ having a negative eigenvalue, as well as its possible identification with some physically relevant system. The fact that these minimal quadratic algebra can be interpreted as a quadratic extension of a Lie subalgebra also constitutes a problem that deserves to be inspected more in detail, possibly in connection with the deformation theory of quadratic algebras \cite{Bra}. It may be observed that, although the computations have been done for the Schr\"odinger algebras, the ansatz is valid in principle for any other non-semisimple Lie algebra possessing a nontrivial Levi decomposition. 

\medskip

The realization-independent approach provides an alternative way to analyze and generate new examples of superintegrable systems, and allows, to a certain extent, to impose the degree of the constants of the motion associated to a given algebraic Hamiltonian. The realization of the Lie algebra 
$\widehat{S}(n)$ considered in this work is merely one among various possible physical realizations, and the question that arises naturally is whether the algebraic approach can be of use in the context of the classification problem of (super)integrable systems into equivalence classes. The algebraic approach also allows to use the framework of representation theory for an effective description of states. Realizations of Lie algebras by vector fields is still a relatively unexplored subject, at least for the case of non-reductive algebras. For the simple case, it is well known that various realizations are deeply connected with different types of special functions \cite{Mill13,Bas82,Bas89}, such as the already mentioned relation between the simple Lie algebra $\mathfrak{sl}(2,\mathbb{R})$ and the Askey scheme for orthogonal polynomials \cite{Koo86}.

\medskip 
Non semisimple Lie algebras, such as the Schr\"odinger and other kinematical algebras, have found applications in different fields, in particular in the context of holography \cite{Ai12,Dov14,Har15}. It may be inferred that the construction of quadratic algebras and algebraic integrals from polynomials in enveloping algebras of non-semisimple Lie algebras may be relevant for other applications. This connection may also point out a alternative scheme to classify quadratic algebras in a more systematic way, however without loosing their deep relation with physical phenomena. Further work in this direction is currently in progress.

\section*{Acknowledgement}
IM was supported by by Australian Research Council Future Fellowship FT180100099. RCS was   supported by
the research grant PID2019-106802GB-I00/AEI/10.13039/501100011033 (AEI/ FEDER, UE).


\begin{thebibliography}{99}

\bibitem{Fre91} L. Freidel, J. M. Maillet, Quadratic algebras and integrable systems, Phys. Lett. A 262, 278--284 (1991) 

\bibitem{Kress07} J. M. Kress, Equivalence of superintegrable systems in two dimensions, Phys. At. Nucl. 70, 560-566 (2007)

\bibitem{Mill13} W. Miller, Jr., S. Post and P. Winternitz, Classical and Quantum Superintegrability with applications, J. Phys. A 46, 423001 (2013)

\bibitem{Yat18} L. A. Yates, P. D. Jarvis, Hidden supersymmetry and quadratic deformations of the space- time conformal superalgebra J. Phys. A: Math. Theor. 51, 145203 (2018)

\bibitem{Lati21} D. Latini, I. Marquette, Y.-Z. Zhang, Embedding of the Racah algebra $R(n)$ and superintegrability, Ann. Phys. 426, 168397 (2021)  

\bibitem{Yi18} Y. Liao, I. Marquette, Y.-Z- Zhang, Quantum superintegrable system with a novel chain structure of quadratic algebras, J. Phys. A. Math. Theor. 51, 255201 (2018) 

\bibitem{cam21}
R. Campoamor-Stursberg, I. Marquette, Hidden symmetry algebra and construction of quadratic algebras of superintegrable systems, Annals of Physics 424, 168378 (2021)

\bibitem{cor21}
F. Correa, M. A. del Olmo, I. Marquette, J. Negro, Polynomial algebras from su(3) and a quadratically superintegrable model on the two sphere, J. PhysA. Math. and Theor 54, 015205 (2021)

\bibitem{Gel} I. M. Gel'fand, Centre of the infinitesimal group ring, Mat. Sbornik  26, 103--112 (1950) 

\bibitem{Dix} J. Dixmier, Sur l'alg\`ebre enveloppante d'une alg\`ebre de Lie nilpotente, Archiv Math. 10, 321--326 (1959)

\bibitem{Tro} V. V. Trofimov, Finite dimensional representations of Lie algebras and integrable systems, Mat. Sb. 3(4), 610--621 (1980)
 
\bibitem{Ni72} U. Niederer, The maximal kinematical invariance group of the free Schr\"odinger equation, Helvetica Physica Acta 45, 802--808 (1972)

\bibitem{Fei04} P. Feinsilver, J. Kocik, B. Schott, Representations of the Schr\"odinger algebra and Appell systems, Fortschritte der Physik 52, 343–359
(2004)

\bibitem{Ai13}
 N. Aizawa, Y. Kimura, J. Segar, Intertwining operators for $\ell$-conformal Galilei algebras and hierarchy of invariant equations, J. Phys. A: Math. Theor. 46, 405204 (2013)

\bibitem{Ai15}
 N. Aizawa, R. Chandrashekar, J. Segar, Lowest weight representations, singular vectors and invariant equations for a class of conformal Galilei algebras, SIGMA 11, 002 (19 pp) (2013)
 
\bibitem{cam19} R. Campoamor-Stursberg, I. Marquette. Generalized conformal pseudo-Galilean algebras and their Casimir operators, J. Phys. A: Math. Theor. 52, 475202 (2019)
 
\bibitem{AMA75} L. Abellanas, L. Mart\'{\i}nez Alonso, A general setting for Casimir invariants, J. Math. Phys. 16, 1580--1584 (1975)
 
\bibitem{Dix1} J. Dixmier, {\it Alg\'ebres enveloppantes}, Paris, Gauthier-Villars 1974

\bibitem{Sn14}
L. \v{S}nobl and P. Winternitz, \textit{Classification and Identification of Lie Algebras}, Rhode Island, 
 Amer. Math. Society (2014)
 
\bibitem{Kam} E. Kamke. {\it Differentialgleichungen. L\"osungsmethoden und L\"osungen II},  Stuttgart, B. G. Teubner 1979

\bibitem{C40} R. Campoamor-Stursberg. Explizite Formeln f\"ur die Casimiroperatoren des semidirekten
Produktes einer Heisenberg Lie-Algebra $\frak{h}$ mit der
einfachen Lie-Algebra $\frak{sl}\left( 2,\mathbb{C}\right)$, Acta
Phys. Polonica B 35, 2059-2069 (2004)

 \bibitem{Abs} M. Abramovich, I. A. Stegun (Eds), {\it Handbook of Mathematical Functions with Formulas, Graphs, and
Mathematical Tables}, New York, Wiley Interscience 1984
 
\bibitem{Bra} A. Braverman, A. Joseph, The minimal realization from deformation theory, J. of Algebra 205, 13--36  (1998)

\bibitem{Bas82} D. Basu, B. Wolf, The unitary irreducible representations of $\mathfrak{sl}(2,\mathbb{R})$ in all subgroup reductions, J. Math. Phys. 23,    189-205 (1982)

\bibitem{Bas89} D. Basu, Representations of $\mathfrak{sl}(2,\mathbb{R})$ in a Hilbert space of analytic functions and a class of integral transforms, J. Math. Phys. 30, 1-8 (1989)

\bibitem{Koo86} T. Koornwinder, {\it Group theoretic interpretations of Askey's scheme of hypergeometric orthogonal polynomials} In: Alfaro M., Dehesa J.S., Marcellan F.J., Rubio de Francia J.L., Vinuesa J. (eds) Orthogonal Polynomials and their Applications. LNM vol 1329, Springer, Berlin, pp 46-72

\bibitem{Ai12}
N. Aizawa and V. K. Dobrev, Schr\"odinger algebra and non-relativistic holography, J. Phys.: Conf. Ser. 343, 012007 (2012)

\bibitem{Dov14}
V. K. Dobrev, Non-relativistic holography: a group-theoretical perspective, Int. J. Mod. Phys. A29, 1430001 (2014)

\bibitem{Har15}
J. Hartong, E. Kiritsis and N. A. Obers, Lifshitz space-times for Schr\"odinger holography  Phys. Lett B 746, 318 (2015)










   
\end{thebibliography}
\end{document}